\documentclass[conference]{IEEEtran}

\usepackage{graphicx}
\usepackage{color}
\usepackage[usenames,dvipsnames,svgnames,table]{xcolor}
\usepackage{multirow}

\DeclareGraphicsExtensions{.eps}

\usepackage{siunitx}

\ifCLASSINFOpdf
\else
\fi

\usepackage[cmex10]{amsmath}
\usepackage{verbatim}
\usepackage[footnotesize, tight]{subfigure}

\usepackage[english]{babel}
\usepackage{blindtext}

\usepackage{cite}

\usepackage{amsmath}

\ifCLASSOPTIONcompsoc
  \usepackage[caption=false,font=normalsize,labelfont=sf,textfont=sf]{subfig}
\else
  \usepackage[caption=false,font=footnotesize]{subfig}
\fi

\DeclareRobustCommand*{\IEEEauthorrefmark}[1]{%
  \raisebox{0pt}[0pt][0pt]{\textsuperscript{\footnotesize\ensuremath{#1}}}}
\usepackage{siunitx}

\begin{document}

%

\title{28 GHz Microcell Measurement Campaign for Residential Environment }

\author{
    \IEEEauthorblockN{C. U. Bas\IEEEauthorrefmark{1}, {\it Student Member, IEEE},
    R. Wang\IEEEauthorrefmark{1}, {\it Student Member, IEEE},
    S. Sangodoyin\IEEEauthorrefmark{1}, {\it Student Member, IEEE}, \\
    S. Hur\IEEEauthorrefmark{3}, {\it Member, IEEE},
    K. Whang\IEEEauthorrefmark{3}, {\it Member, IEEE},
    J. Park\IEEEauthorrefmark{3}, {\it Member, IEEE}, \\
    J. Zhang\IEEEauthorrefmark{2}, {\it Fellow, IEEE}, 
    A. F. Molisch\IEEEauthorrefmark{1}, {\it Fellow, IEEE}  
    }\\
    \IEEEauthorblockA{\IEEEauthorrefmark{1}University of Southern California, Los Angeles, CA, USA,}
    
    \IEEEauthorblockA{\IEEEauthorrefmark{2}Samsung Research America, Richardson, TX, USA}
    
     \IEEEauthorblockA{\IEEEauthorrefmark{3}Samsung Electronics, Suwon, Korea}
}


\maketitle

\begin{abstract}
This paper presents results from the (to our knowledge) first double-directionally resolved measurement campaign at mm-wave frequencies in a suburban microcell. The measurements are performed with a real-time channel sounder equipped with phased antenna arrays that allows electrical beam steering in microseconds, and which can measure path-loss of up to 169 dB. 
Exploiting the phase coherency of the measurements in the different beams, we obtain both directional and omnidirectional channel power delay profiles without any delay uncertainty. We present statistics of channel characteristics such as path-loss, shadowing and delay spread results for line-of-sight and non-line-of-sight cases, as well as sample results for power angular spectrum and extracted multi-path components. 

\end{abstract}

%
\IEEEpeerreviewmaketitle

\section{Introduction}

Due to the ever-increasing demand for wireless data, current networks are becoming overburdened. While a variety of different techniques will be used to alleviate this congestion and enable future growth \cite{Falahy_2017_tech} \cite{andrews2014will}, making new spectrum available is among the most promising approaches. For this reason, there is great interest in developing wireless communications systems in the frequency spectrum beyond \SI{6}{GHz}, which up to now has been mostly fallow \cite{Boccardi_2014_five}. In a recent ruling, the frequency regulator in the USA, the Federal Communications Committee, has allowed usage of more than 10 GHz of bandwidth of that frequency range for new services - considerably more than currently used in all wireless services taken together. Other countries are expected to follow suit, and  frequency bands such as \SI{28}{GHz}, \SI{38}{GHz}, \SI{60}{GHz} and \SI{75}{GHz} are being considered for fifth generation (5G) cellular networks \cite{federal_2015_matter}. For outdoor applications the 28 GHz currently enjoys the greatest interest, since the comparatively low frequency (compared to other mm-wave bands) allows a lower-cost implementation of many components. 

The design and deployment planning of any wireless system requires a thorough understanding of the wireless propagation channel. For example, path-loss and shadowing characteristics determine distance-dependent outage probability, while delay dispersion determines spacing of subcarriers (in OFDM) or length of equalizers (in single-carrier systems). At the same time, it must be remembered that the channel characteristics strongly depend on the propagation environment. Thus, measurements of channel characteristics and creation of models derived from them {\em in the environment of interest} are vital. While there have been several measurement campaigns for channel characteristics in {\em urban} microcellular environments (see below), to the best of our knowledge no such measurements exist in {\em suburban} microcell environments. The current paper aims to close this gap. In particular, we will present results from a measurement campaign in a microcellular, suburban environment with a directionally resolving, wide-band channel sounder, and extract some key channel characteristics.
 
{\bf Existing work:} As mentioned above, a number of directionally resolved measurements have been performed in urban microcellular environments\footnote{Due to space restrictions, we do not review non-directionally resolved measurement campaigns here}. The works in \cite{Rappaport_et_al_2015_TCom,maccartney2014omnidirectional,sulyman2014radio} in downtown New York City. Refs. \cite{park2016millimeter} and \cite{Hur_et_al_2015_EuCAP} provide results in various cities in Korea, while  \cite{zhao2017channel} reports channel measurements conducted at \SI{32}{GHz} on a University campus in Beijing, China. All these environments are densely built up, with high-rise buildings ($\ge 5$ floors and/or contiguous facades). All of the measurements use mechanically rotating horn antennas to extract the directional characteristics and make use of the high antenna gain to improve the link budget; the drawback of this approach is that it is very time-intensive, often requiring hours to scan a single measurement location, and thus naturally limiting the number of locations underlying the measurements. 

The suburban measurements reported in the literature have been mainly focused on Fixed Wireless Access (FWA, also known as LMDS) systems \cite{Papazian_1997_study,Eldeen_2010_modelling,Xu_2000_Measurements}, i.e both transmit and receive antennas are above rooftops, and thus different from the scenario considered here. Similarly, measurements in suburban environments at carrier frequencies below 6 GHz in suburban microcells exist, but due to the different frequency range cannot provide any information about mm-wave propagation.

{\bf Contributions:}
In this work, we present the results from the first \SI{28}{GHz} channel sounding campaign in a residential suburban cellular scenario with directionally resolvable results for both TX and RX. The measurements are performed with a real-time channel sounder equipped with phased antenna arrays \cite{bas_2017_realtime}. The phased arrays form beams at the different TX and RX angles and switch between these beams in microseconds, which allows measurement at a large number of locations within reasonable time, and ensures minimal variation in the environment during the performing of the measurements. We provide key channel characteristics, such as path-loss, shadowing, and delay spread, for both line-of-sight (LoS) and non-line-of-sight (NLoS) situations and present sample results for power angular spectrum and extracted multi-path components.

The rest of the paper is organized as follows. Section \ref{sec:meas} discusses the channel sounder setup and the measurement environment.  Section \ref{sec:data} explains the data processing. Section \ref{sec:results}  presents measurement results. Finally Section \ref{sec:conc} summarizes results and suggests directions for future work.

  \begin{table}[tbp]\centering
  \caption{Sounder specifications}
  \renewcommand{\arraystretch}{1.1}
\begin{tabular}{l|c}
    \hline
    \multicolumn{2}{c}{Hardware Specifications} \\ \hline \hline
    Center Frequency & 27.85 GHz\\
    Instantaneous Bandwidth & 400 MHz\\
    Antenna array size & 8 by 2 (for both TX and RX) \\
    Horizontal beam steering & -45 to 45 degree \\
    Horizontal 3dB beam width & 12 degrees\\
    Vertical beam steering & -30 to 30 degree \\
    Vertical 3dB beam width & 22 degrees\\
    Horizontal/Vertical steering steps & 5 degrees\\
    Beam switching speed & 2$\mu s$ \\
    TX EIRP & 57 dBm \\
    RX noise figure & $\le$ 5 dB \\ 
    ADC/AWG resolution & 10/15-bit \\
    Data streaming speed & 700 MBps \\ \hline
    \multicolumn{2}{c}{Sounding Waveform Specifications} \\ \hline \hline
    Waveform duration & 2 $\mu s$ \\
    Repetition per beam pair & 10 \\
    Number of tones & 801 \\
    Tone spacing & 500 kHz \\
    PAPR & 0.4 dB \\ 
    Total sweep time & 14.44 ms\\ \hline      
  \end{tabular} \label{specs}
\end{table}

\begin{figure}[tbp]\centering
  \includegraphics[width=0.5\linewidth]{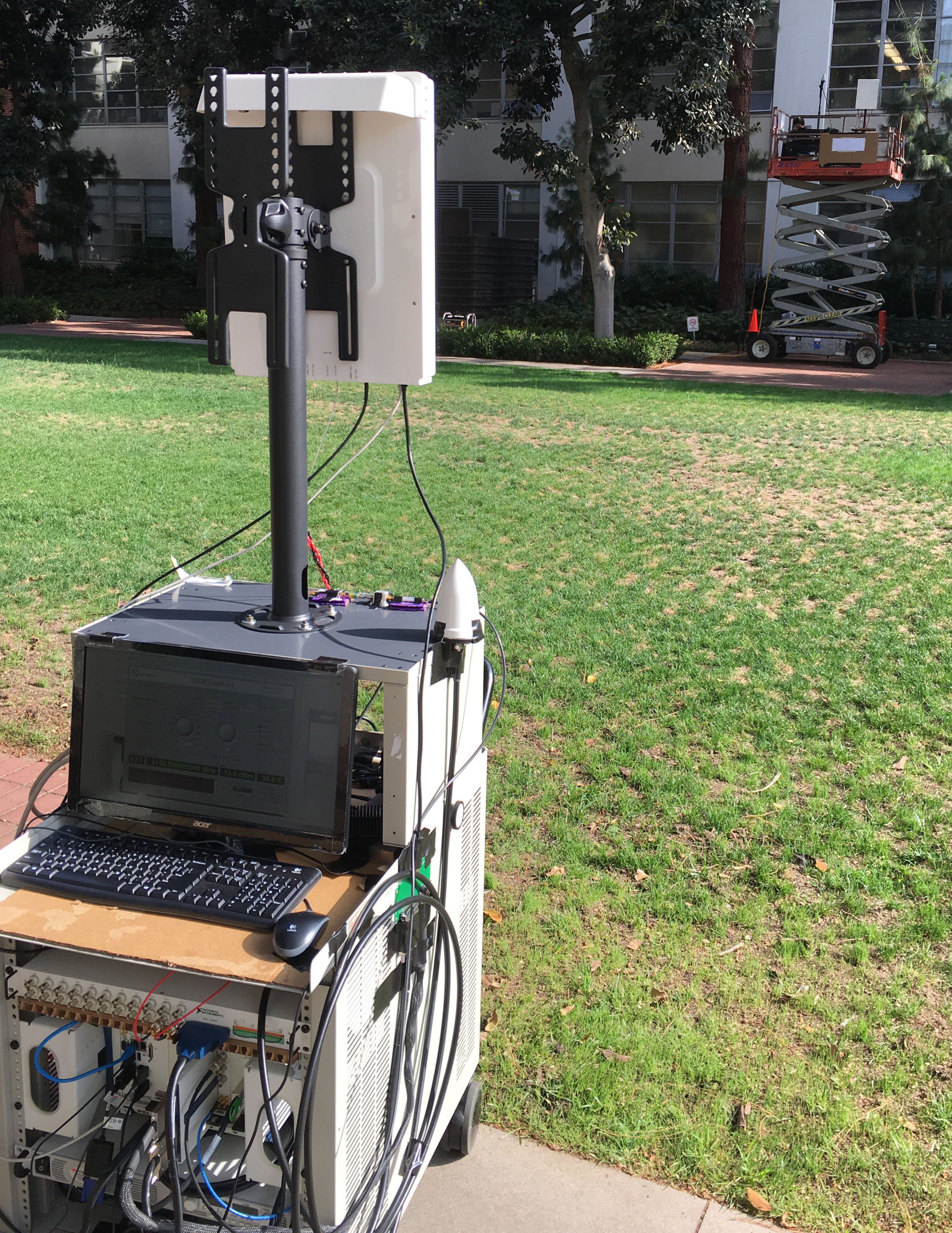}\caption{RX setup}\label{Fig:rx_tx.JPG}
\end{figure}

\section{Measurement Campaign} \label{sec:meas}

\subsection{Channel Sounder Setup}
In this campaign, we used a switched-beam, wide-band mm-wave sounder with 400 MHz real-time bandwidth \cite{bas_2017_realtime}.The sounding signal is a multi-tone signal which consists of equally spaced 801 tones covering 400 MHz. A low peak to average power ratio (PAPR) of $0.4$ dB is achieved by manipulating the phases of individual tones as suggested in \cite{Friese1997multitone}. This allows us to transmit with power as close as possible to the 1 dB compression point of the power amplifiers without driving them into saturation.  

Both the TX and the RX have phase arrays capable of forming beams which can be electronically steered with $5^{\circ}$ resolution in the range of $[-45^{\circ}, 45^{\circ}]$ in azimuth and $[-30^{\circ}, 30^{\circ}]$ in elevation. This decreases the measurement time for one RX location from hours to milliseconds. During this measurement campaign we only utilize a single elevation angle $0^{\circ}$ with 19 azimuth angles both for the TX and the RX. With an averaging factor of 10, the total sweep time is \SI{14.44}{ms}(without averaging it can be as low as \SI{1.444}{ms}) for 361 total beam pairs. Since phased arrays cover $90^{\circ}$ sectors, we rotated the RX to $\{0^{\circ},90^{\circ},180^{\circ},270^{\circ}\}$ to cover $360^{\circ}$ while using a single orientation at the TX. Consequently, for each measurement location, we obtain a frequency response matrix of size 19 by 72 by 801.  Moreover, thanks to the beam-forming gain, the TX EIRP is 57 dBm, and the measurable path loss is 159 dB without considering any averaging or processing gain. By using GPS-disciplined Rubidium frequency references, we were able to achieve both short-time and long-time phase stability. Combined with the short measurement time this limits the phase drift between TX and RX, enabling phase-coherent sounding of all beam pairs even when TX and RX are physically separated and have no cabled connection for synchronization. Consequently, the directional power delay profiles (PDP) can be combined easily to acquire the omnidirectional PDP. Table \ref{specs} summarizes the detailed specification of the sounder and the sounding waveform. Ref. 
 \cite{bas_2017_realtime} discusses further details of the sounder and the validation measurements.

\begin{figure*}[tbp]\centering
  \includegraphics[width=0.85\linewidth]{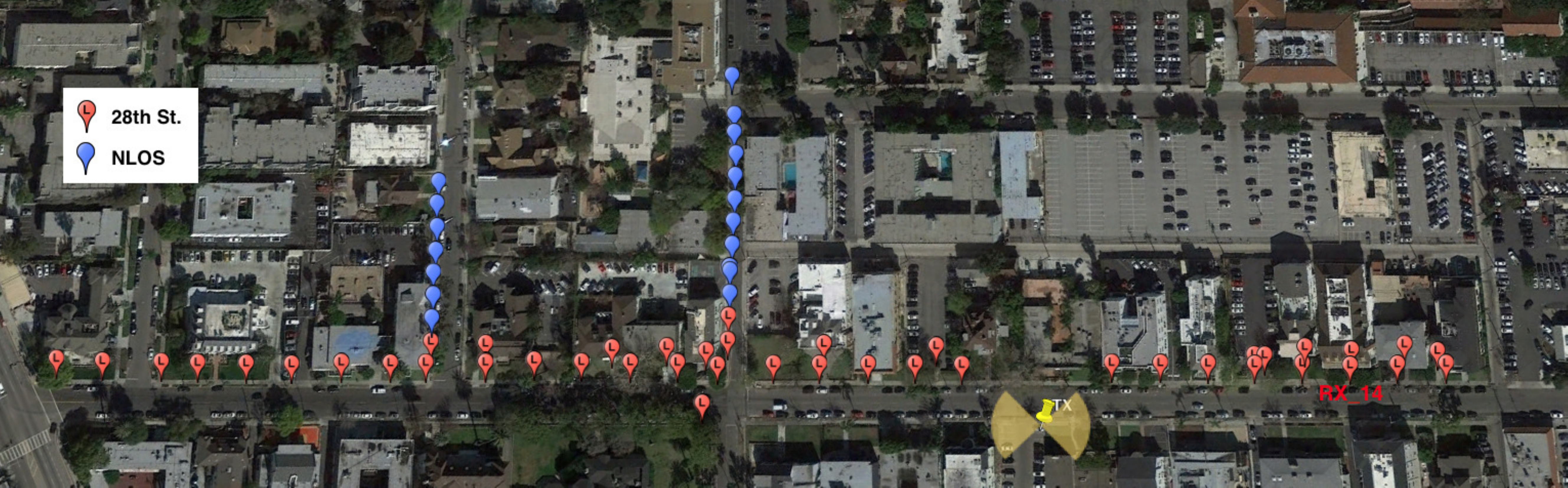}\caption{Measurement locations}\label{Fig:locs}
\end{figure*}

\subsection{Measurement Environment}
\begin{figure}[tbp]\centering
  \includegraphics[width=0.8\linewidth]{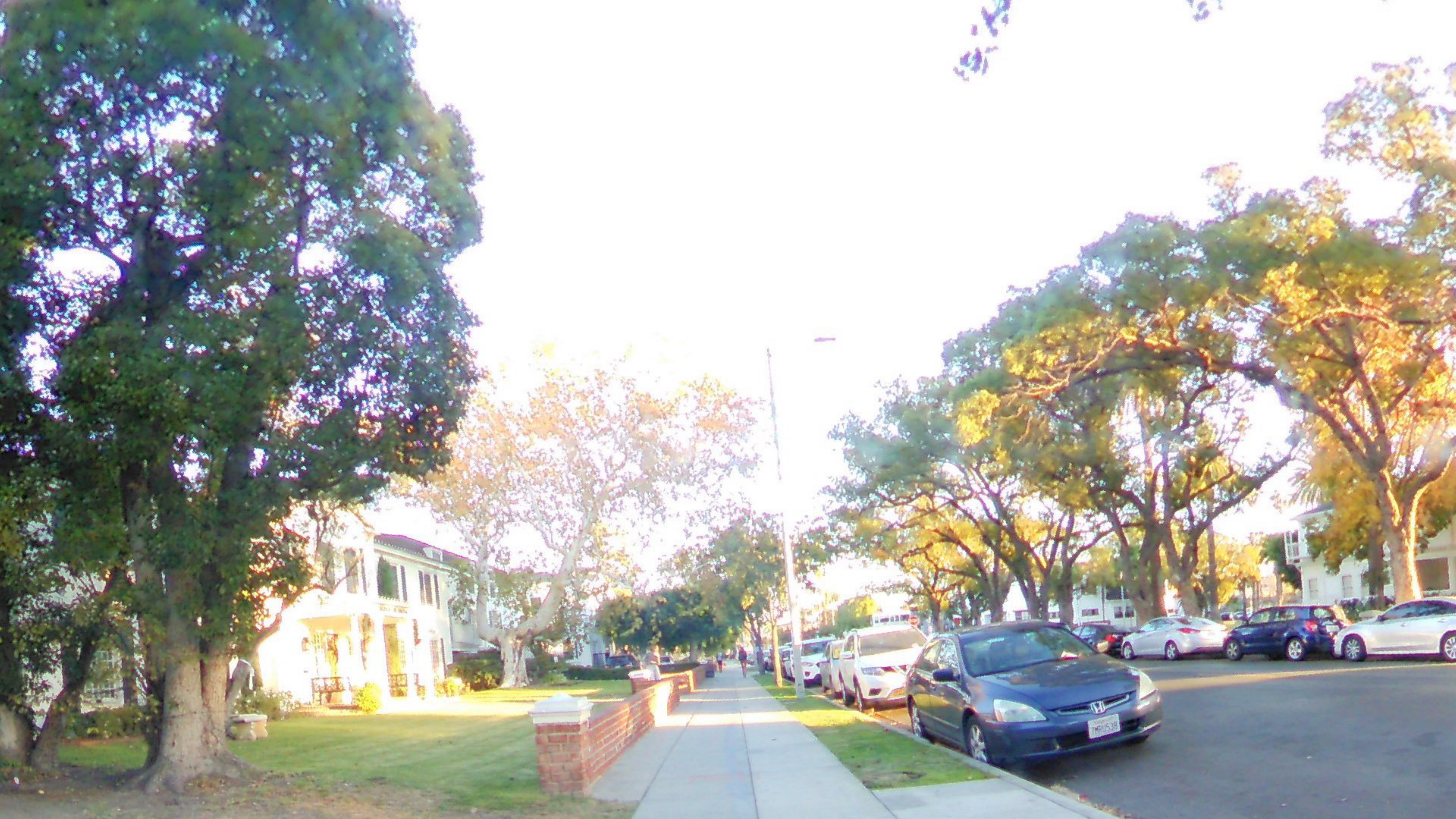}\caption{RX View on 28th St facing east}\label{Fig:rx_view}
\end{figure}

\begin{figure}[tbp]\centering
  \includegraphics[width=0.8\linewidth]{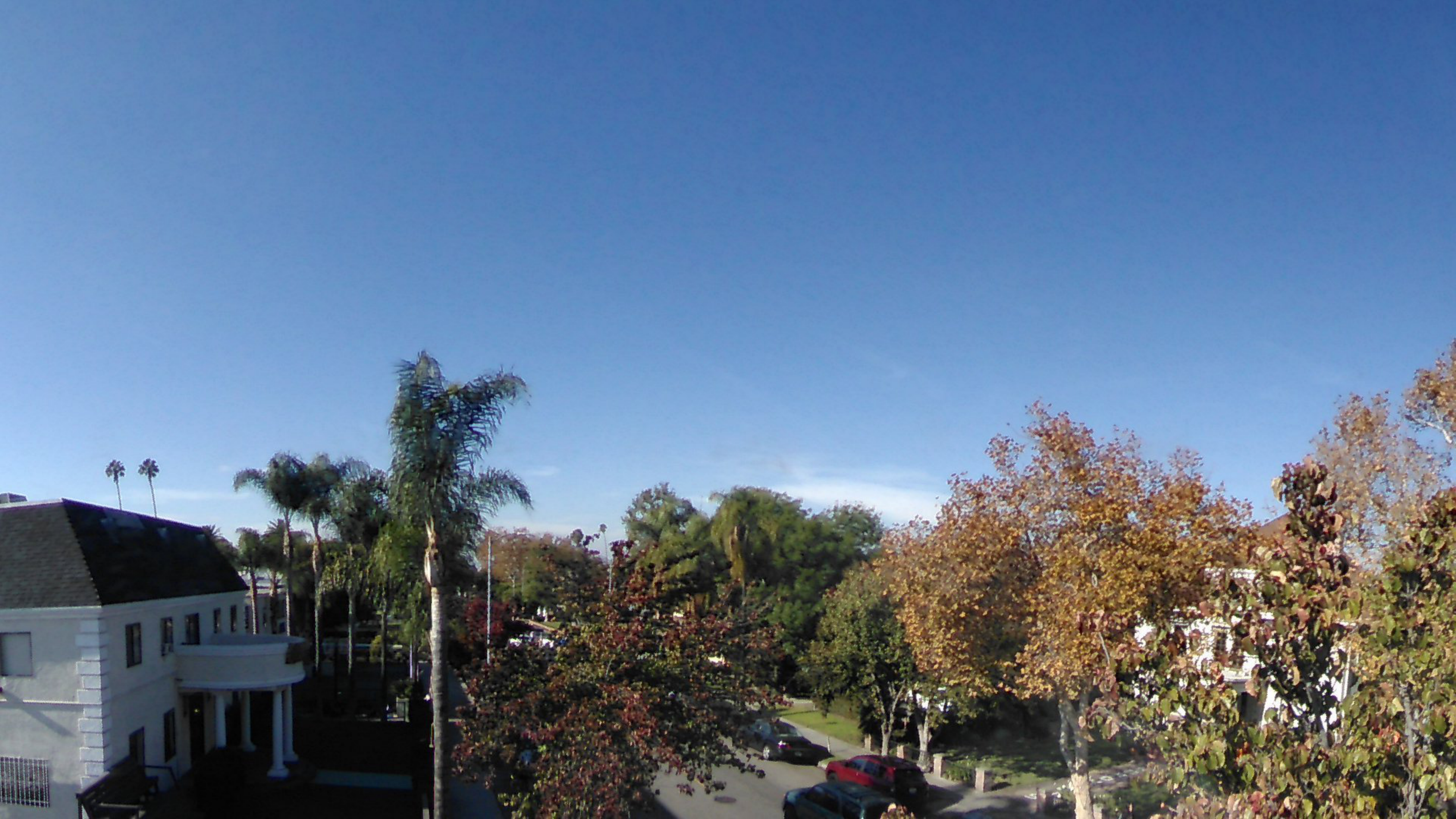}\caption{TX View on 28th St facing west}\label{Fig:tx_view}
\end{figure}

The measurements were performed in a typical US suburban residential area at/near 28th Street in Los Angeles, CA, USA\footnote{note that despite the location in Los Angeles, the building height and density is suburban, not metropolitan, as can also be seen from Figures \ref{Fig:locs},\ref{Fig:rx_view} and \ref{Fig:tx_view}} populated with 2 to 3 story houses along a street that contains trees and other foliage, see Figure \ref{Fig:locs}. Consequently, the measurements mimic a real-life cellular deployment scenario including the effect of foliage penetration loss. To imitate a microcell scenario, the TX is placed on a scissor lift at the height of \SI{7.5}{m} while the RX is on a cart, and the RX antenna height is \SI{1.8}{m}. The bore-sight of the $90^{\circ}$ TX sector is parallel to the 28th St and faced towards to RX.

The RX locations are chosen for 2 different scenarios. In the first one, the RX is placed on the same street (28th St) with the TX. Since in some cases the direct path is blocked by foliage or other surrounding objects, throughout the paper we call this data 28th St. instead of LOS. In the second scenario, the RX is located on the two crossing streets to create NLOS links. All RX locations are either on the sidewalks or in the front yards of the surrounding houses. The range of TX-RX separation vary from \SI{36}{m} to \SI{400}{m} for the 28th St and from \SI{130}{m} to \SI{273}{m} for NLOS.


\begin{figure}[tbp]\centering
  \includegraphics[width=0.85\linewidth]{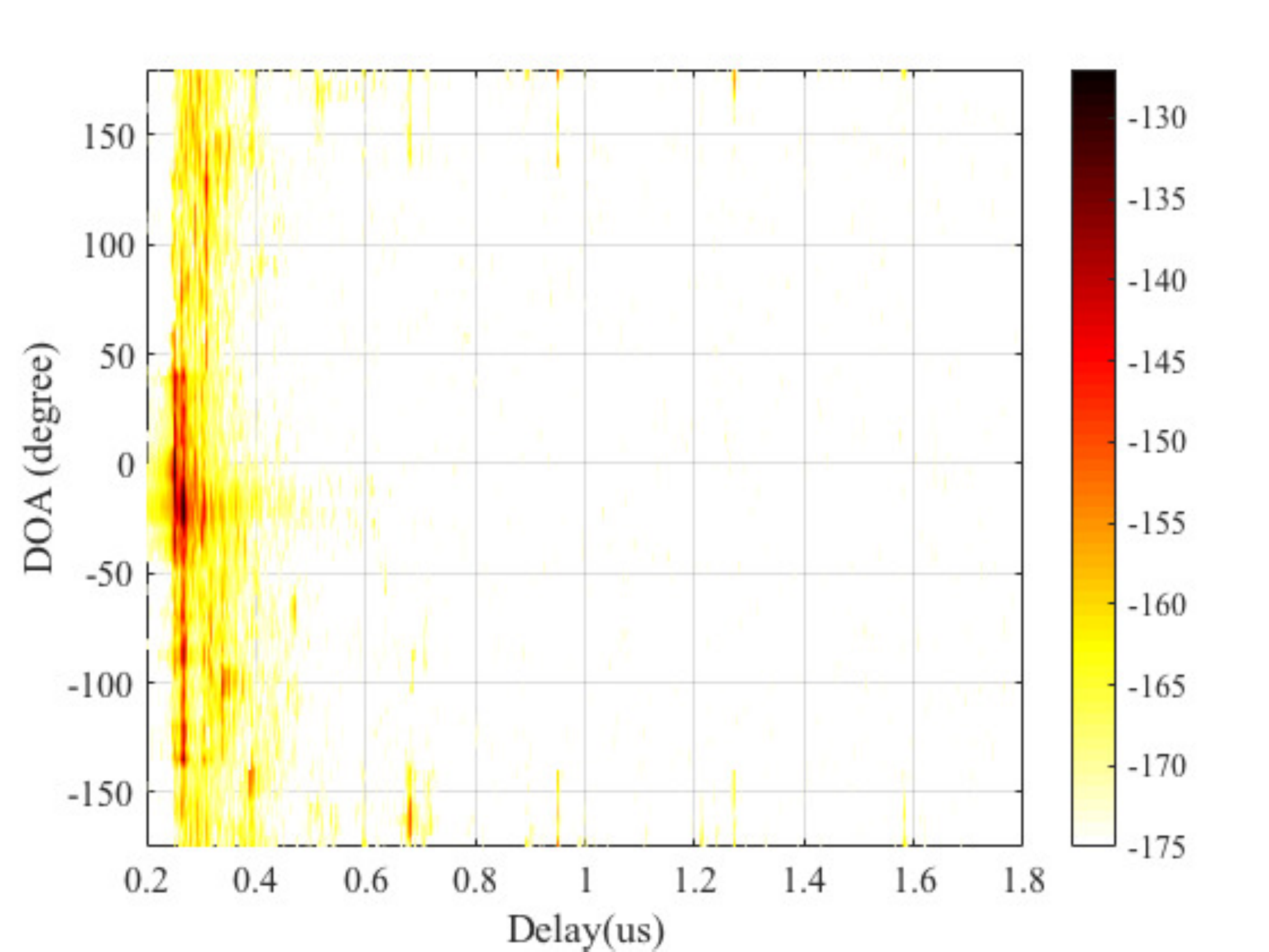}\caption{Power-Angular Delay Profile for RX\_14}\label{Fig:padp}
\end{figure}

\begin{figure*}[htbp]\centering
  \includegraphics[width=0.85\linewidth]{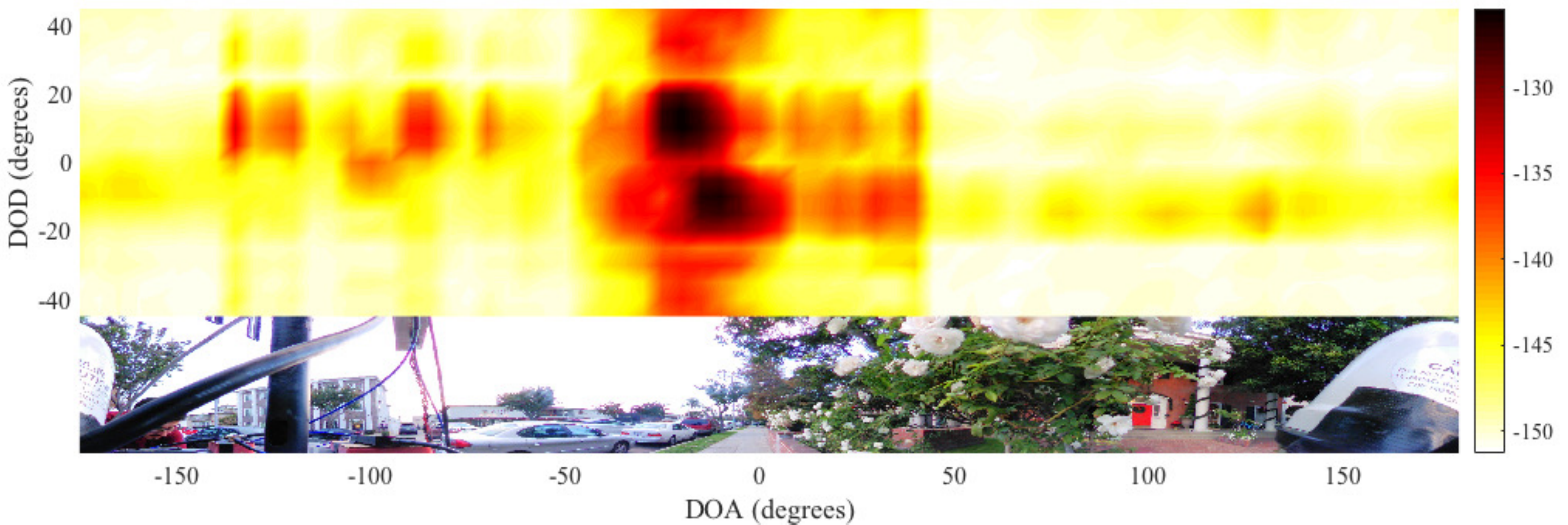}\caption{Power-Angular Spectrum for RX\_14}\label{Fig:pas}
\end{figure*}

\begin{figure}[tbp]\centering
  \includegraphics[width=0.8\linewidth, viewport=38 180 570 600, clip=true]{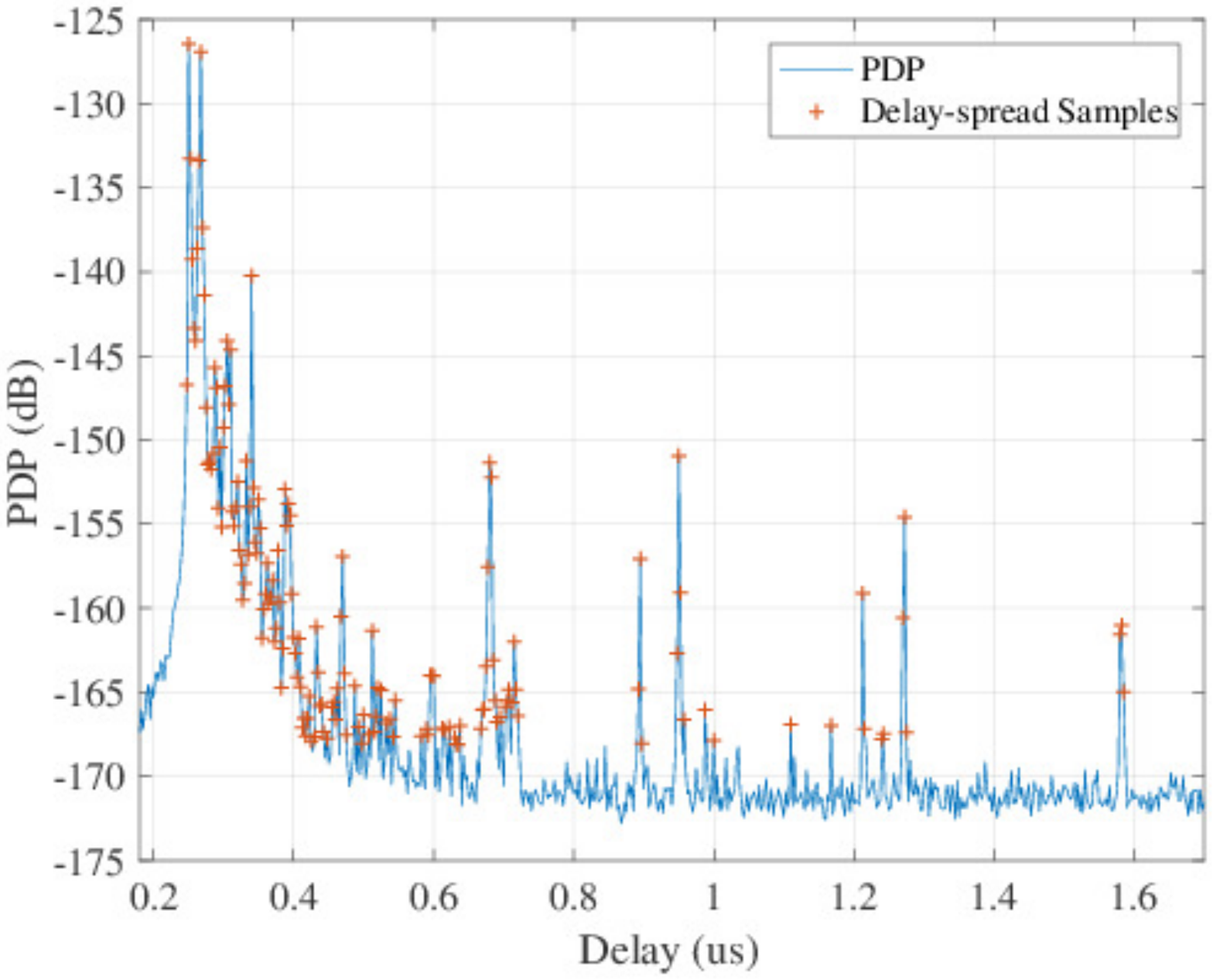}\caption{Power Delay Profile  for RX\_14}\label{Fig:pdp_14}
\end{figure}

\section{Data Evaluations} \label{sec:data}

The directional power delay profile (PDP) for the TX beam and RX beam with the azimuth angles  $\theta_{TX}$  and  $\theta_{RX}$  is estimated as;
\begin{equation}
  \resizebox{.9 \linewidth}{!} 
{
$   P(\theta_{TX},\theta_{RX},\tau) = \bigg\vert \mathcal{F}^{-1} \left\{ H_{\theta_{TX},\theta_{RX}}\left(\vec{f} \right) ./ H_{cal}\left(\vec{f} \right) \right\} \bigg\vert ^2$
}
\end{equation}
where $\theta_{RX}\in[-175,180]$,  $\theta_{TX}\in[-45,45]$, $\mathcal{F}^{-1}$ denotes inverse Fourier transform, $H_{\theta_{TX},\theta_{RX}}(\vec{f})$ and $H_{cal}(\vec{f})$ are the frequency responses for TX beam $\theta_{TX}$ and RX beam $\theta_{RX}$ and, the calibration response respectively; $\vec{f}$ are the used frequency tones, and $./$ is element-wise division. 

Then the angular power spectrum can be calculated as:
\begin{equation}
 {\displaystyle  PAS(\theta_{TX},\theta_{RX}) = \sum_{\tau} P(\theta_{TX},\theta_{RX},\tau)}
\end{equation}
Figure \ref{Fig:pas} shows the $PAS(\theta_{TX},\theta_{RX})$ for RX\_14 (marked in Figure \ref{Fig:locs}) along with the RX view.

Furthermore, similar to \cite{hur_synchronous_2014}, the omni-directional power delay profile ($PDP$) power angular-delay profiles for RX and TX ($PADP_{RX/TX}$) are calculated as follows. 
\begin{align} 
  PDP (\tau) &= {\displaystyle \max_{\theta_{TX}} \max_{\theta_{RX}}  P(\theta_{TX},\theta_{RX},\tau)} \label{eq:pdp}\\ 
  PADP_{RX}(\theta_{RX},\tau) &= {\displaystyle \max_{\theta_{TX}} P(\theta_{TX},\theta_{RX},\tau)} \\
  PADP_{TX}(\theta_{TX},\tau) &= {\displaystyle \max_{\theta_{RX}} P(\theta_{TX},\theta_{RX},\tau)}
 \end{align} 
For the same RX location, Figures \ref{Fig:padp} and \ref{Fig:pdp_14} show the $PADP_{RX}$ and $PDP$ respectively. Additionally, we also consider directional PDP which is defined as the PDP acquired from the TX-RX beam pair with the highest received power. Finally, the path-loss $P_{RX}$ is given by:
\begin{equation}
 {\displaystyle  P_{RX} =  \sum_{\tau} PDP(\tau)} \label{eq:pl}
\end{equation}

\section{Results} \label{sec:results}

\begin{figure}[tbp]\centering
  \includegraphics[width=0.8\linewidth, viewport=38 180 570 600, clip=true]{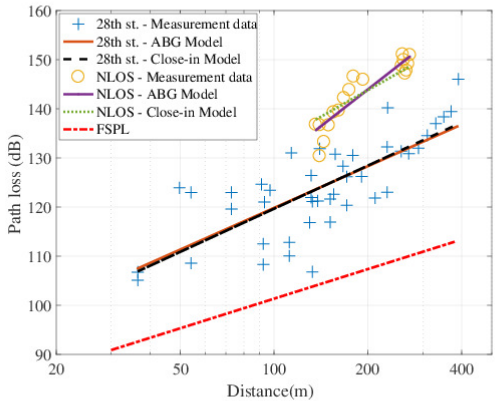}\caption{Path-loss for 28th St. and NLoS locations}\label{Fig:pl}
\end{figure}

\subsection{Path Loss}
There are two approaches commonly used for path loss modeling in mm-wave; alpha-beta-gamma (ABG) and close-in (CI) models \cite{Rappaport_2015_Wideband} \cite{Molisch_2016_eucap}. For a single frequency band, both can be simplified into:

\begin{equation} 
  PL(d)= 10nlog_{10}\left(d/1m \right) + P_0 + \chi_\sigma
\end{equation}
However, they differ in the estimation of model parameters $n$ and $P_0$. For the CI method, $P_0$ is given by $20log_{10}(4\pi f/c)$ which is the free-space path loss for 1m TX-RX separation at the frequency $f$ where $c$ is the speed of light. Once the $P_0$ is fixed, the path-loss exponent (PLE) $n$ is estimated from measurement data via minimum mean square error estimation. In ABG method, both $P_0$ and $n$ are estimated together from measurement data with least-squares regression. For both models $\chi_\sigma$ is a Gaussian random variable with 0 mean and standard deviation of $\sigma$ in dB \cite{Rappaport_2015_Wideband}. 

We used both ABG and CI models to characterize the path-loss for both omnidirectional and directional cases. Figure \ref{Fig:pl} shows the path-loss values for the 28th St and NLoS measurement points along with the ABG and CI fits for the omnidirectional RX, and the theoretical free space path-loss (FSPL). The path-loss model parameters for both directional and omnidirectional RX are given in Table \ref{tab:pl}. For the 28th St, the parameters for the ABG and the CI models are very similar while they differ significantly for NLoS data. Note, however, that while the parameters are different, the resulting line fits {\em in the range of interest}, i.e., the range over which measurements have been made and thus the model is applicable, are quite similar. Table \ref{tab:pl} also summarizes the path-loss models for the directional PDP. In the case of directional PDP, the path-loss components are slightly higher than the omnidirectional case. For 28th St, this is expected, since as the RX moves away from the probability of having an optical LoS decreases, resulting relatively higher attenuation at large distances. 

\begin{figure}[tbp]\centering
\begin{minipage}{0.48\linewidth} \centering
    \includegraphics[width=0.88\linewidth, viewport=130 110 460 680, clip=true]{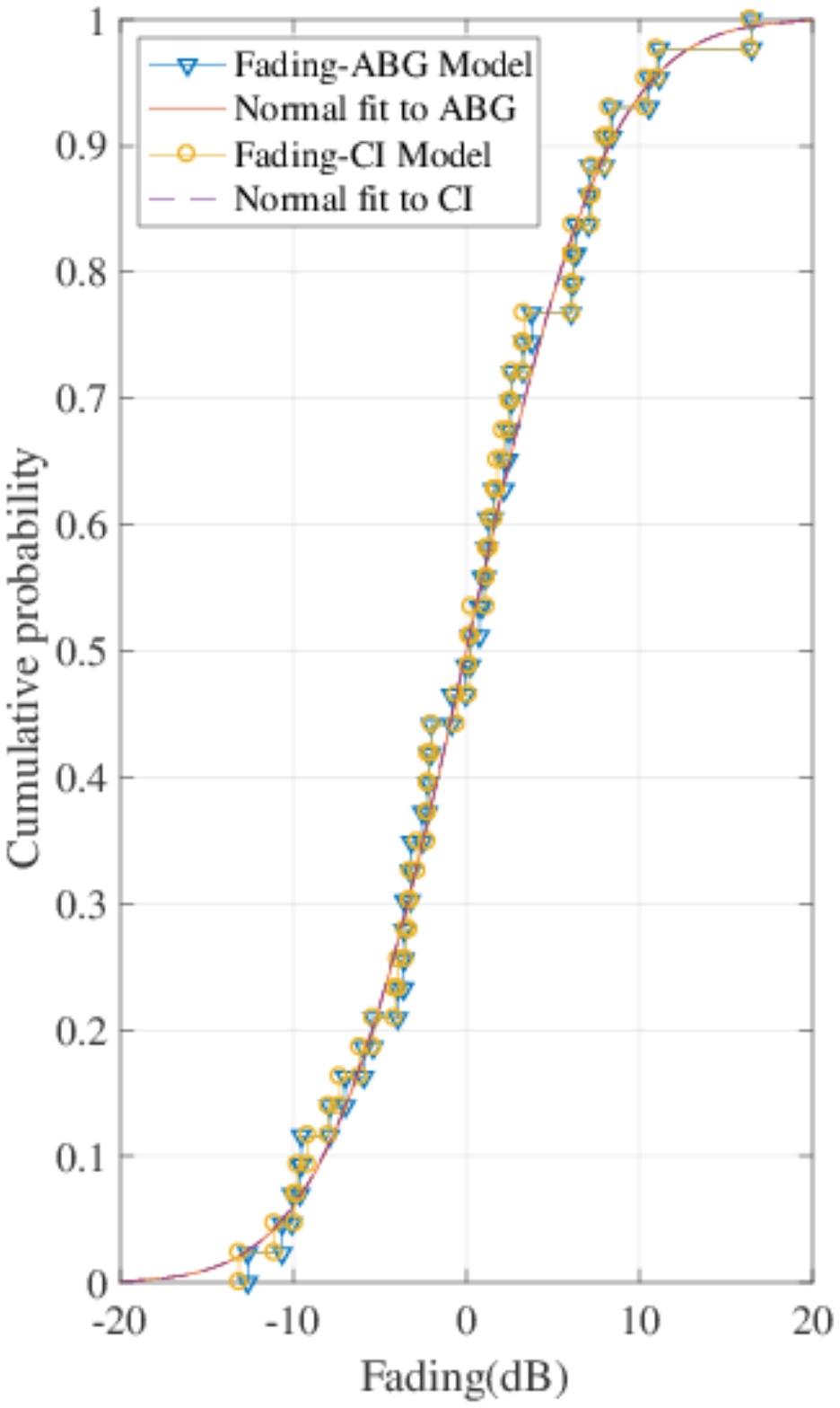}\caption{CDF of the fading for 28th St.}\label{Fig:fadingLOS}
\end{minipage} 
\begin{minipage}{0.03\linewidth} \centering
\end{minipage}
\begin{minipage}{0.48\linewidth} \centering
     \includegraphics[width=0.88\linewidth, viewport=130 110 460 680, clip=true]{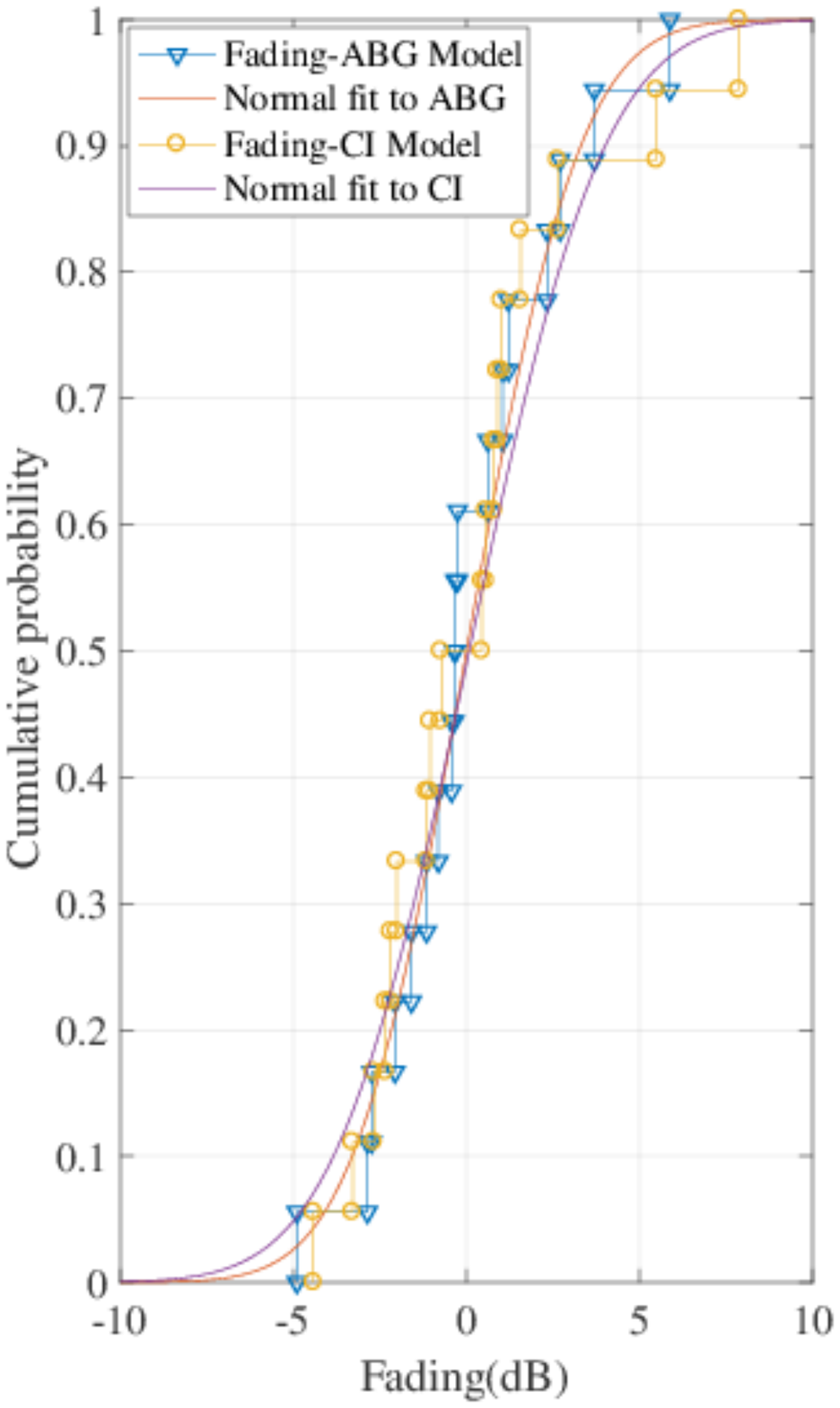}\caption{CDF of the fading for NLoS locations}\label{Fig:fadingNLOS}
  \end{minipage}
\end{figure}

For both omnidirectional scenarios, the cumulative distribution functions (CDF) of shadow fading are given in Figures \ref{Fig:fadingLOS} and \ref{Fig:fadingNLOS}. Both CI and ABG models follow zero-mean Gaussian distributions with the standard deviations listed in Table \ref{tab:pl}. In 28th St, we observe a high shadow fading variance due to the foliage penetration loss and other objects along the street. Furthermore, Table \ref{tab:pl} also shows the P-values of the fits, acquired via Kolmogorov-Smirnov (KS) test which uses the measure of maximum difference between the CDF of the empirical and the hypothetical distributions \cite{massey_1951_kolmogorov}. In all cases KS-test do not reject the hypothetical Gaussian distribution with a P-value larger than 0.7.

 \begin{table}[tbp]\centering \caption{Parameters of the path loss models} \label{tab:pl}
  \scriptsize  
 \begin{tabular}{|l|l|c|c|c|c|c}\hline
     &\multirow{ 2}{*}{ Data }& \multirow{ 2}{*}{n}& \multirow{ 2}{*}{$P_0$} & \multicolumn{2}{c|}{$\chi_\sigma$ } \\ \cline{5-6}
     &&&& $\sigma$ & P-value\\ \hline
     \multirow{ 4}{*}{\rotatebox[origin=c]{90}{omni}} & 28th St - ABG Model & 2.82 & 63.47 & 6.44 & 0.975\\
     &28th St - CI Model & 2.92 & 61.34 & 6.45 &  0.978 \\
     &NLoS - ABG Model & 4.97 & 29.53 & 2.58 &    0.745\\
     &NLoS - CI Model & 3.58 & 61.34 & 3.06 &   0.706 \\      \hline
      \multirow{ 4}{*}{\rotatebox[origin=c]{90}{directional}}&28th St - ABG Model & 3.17 & 58.01 & 7.75  & 0.840\\
     &28th St - CI Model & 3.15 & 61.34 & 7.76    & 0.928 \\
     &NLoS - ABG Model & 5.85 & 18.12 & 4.53 &   0.856 \\
     &NLoS - CI Model & 3.96 & 61.34 & 5.06 & 0.958 \\ 
     \hline   
   \end{tabular}
 \end{table}



\subsection{RMS Delay Spread}

As is common in the literature, we characterize the delay dispersion by the root-mean-square delay spread (RMS-DS), i.e., the second central moment of the power delay profile. 

\begin{equation}
 {\displaystyle S_\tau=\sqrt{\dfrac{\sum\limits_{\hat \tau} PDP(\hat \tau) \hat \tau^2 }{P_{RX}} - \left( \dfrac{\sum\limits_{\hat \tau} PDP( \hat \tau) \hat \tau }{P_{RX}} \right)^2}}
\end{equation}  
where $\left\{ \hat \tau = \tau | PDP(\tau) > 2\sigma_{noise}^2\right\}$ and $\sigma_{noise}^2$ is the noise power for omnidirectional PDP. 

Prior to the RMS-DS calculation, we apply noise filtering to avoid any contribution of the noise floor, which can significantly distort delay spread calculations by creating nonphysical contributions at large delays. Due to the automatic gain control implemented at the RX, the noise level might vary between directional PDPs for different beam pairs. Hence, we first obtain noise-filtered directional PDPs by:
\begin{equation}
\resizebox{1 \linewidth}{!} 
{
    $ P(\theta_{TX},\theta_{RX},\tau)= \begin{cases}
  P(\theta_{TX},\theta_{RX},\tau) & \text{if } P(\theta_{TX},\theta_{RX},\tau) > 4\sigma^2(\theta_{TX},\theta_{RX}) \\
  0 & otherwise
  \end{cases} $
}
\end{equation}
where $\sigma^2(\theta_{TX},\theta_{RX})$ is the noise power for TX beam $\theta_{TX}$ and RX beam $\theta_{RX}$. Figure \ref{Fig:padp} shows the noise-filtered $PADP_{RX}$ for RX\_14. Then the omnidirectional PDP is calculated by using Equation \ref{eq:pdp}.

\begin{figure}[tbp]\centering
\begin{minipage}{0.48\linewidth} \centering
    \includegraphics[width=0.88\linewidth, viewport=130 110 460 680, clip=true]{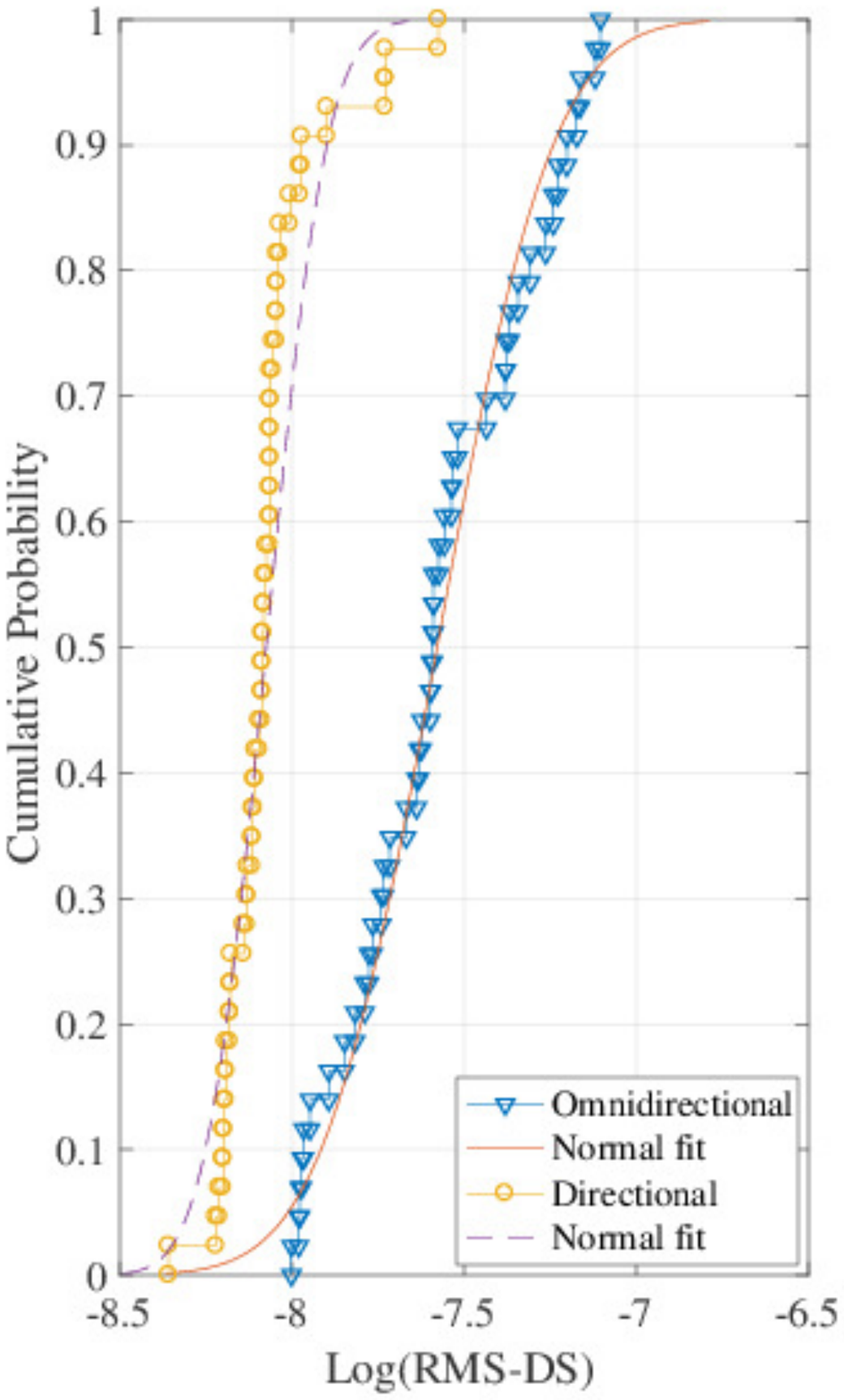}\caption{CDF of the logarithm of RMS-DS for 28th St.}\label{Fig:rms_LOS}
\end{minipage} 
\begin{minipage}{0.03\linewidth} \centering
\end{minipage}
\begin{minipage}{0.48\linewidth} \centering
     \includegraphics[width=0.88\linewidth, viewport=130 110 460 680, clip=true]{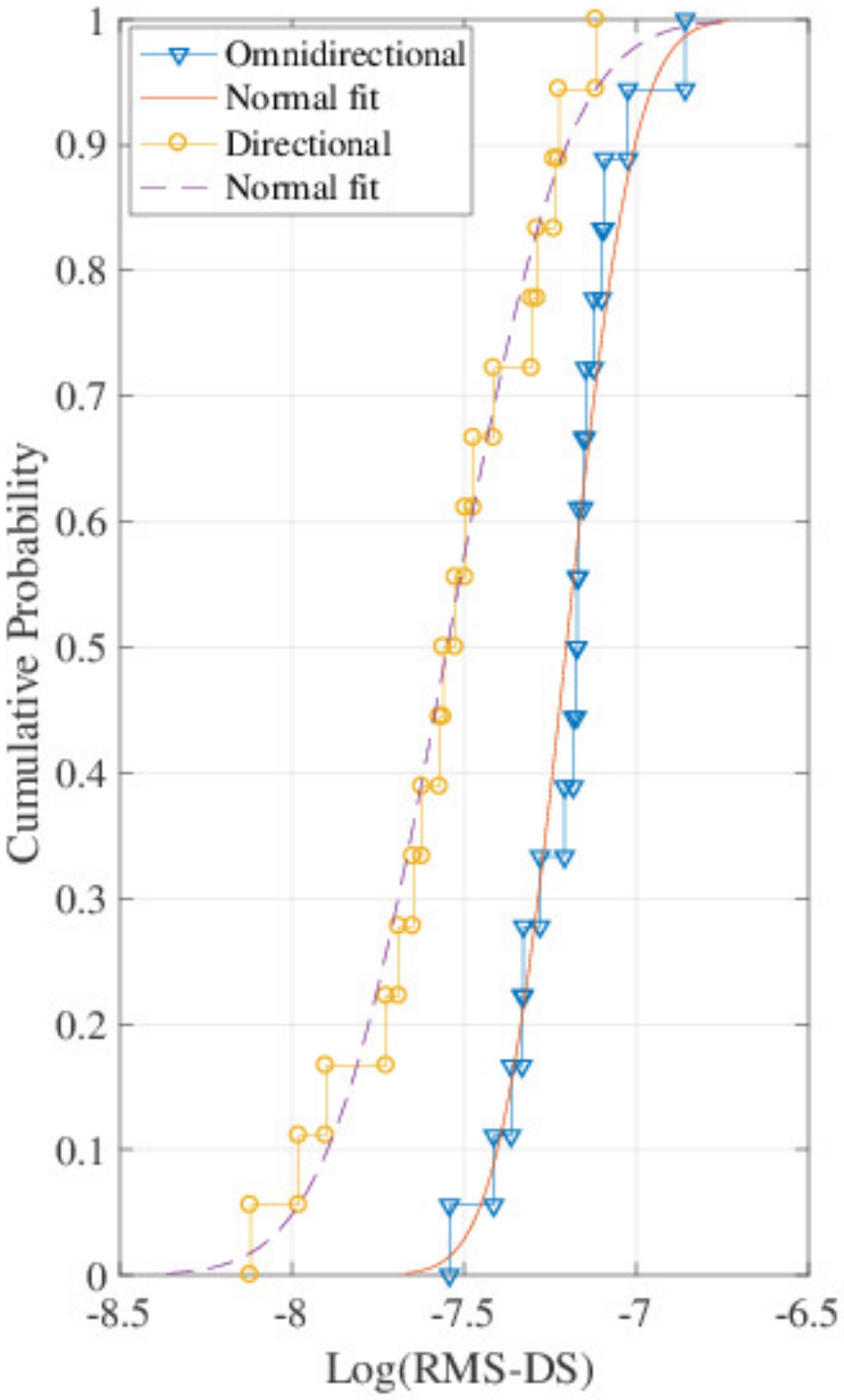}\caption{CDF of the logarithm of RMS-DS for NLoS locations}\label{Fig:rms_NLOS}
  \end{minipage}
\end{figure}

A sample omnidirectional PDP and the samples used for delay spread calculation are shown in Figure \ref{Fig:pdp_14}. Figure \ref{Fig:rms_LOS} and \ref{Fig:rms_NLOS}  show the cumulative distribution functions of the Log(RMS-DS) along with the corresponding Gaussian fits for 28th St and NLOS, respectively. As listed in Table \ref{tab:rms}, the median RMS-DS are \SI{25.63}{ns} for 28th St. and \SI{67.18}{ns} for NLoS. The mean $\mu$ of the Gaussian fits are -7.58 for 28th St and -7.2 for NLOS. In \cite{3GPP_5G_2016}, for urban-micro scenario, they modeled the $\mu$ of the Log(RMS-DS) as $\mu =-0.2Log(1+f)-7.2$ and $\mu =-0.21Log(1+f)-6.88$ for LOS and NLOS respectively. At \SI{27.85}{GHz}, the corresponding $\mu$ values are -7.49 and -7.19 which are well-aligned with our results. We also investigate the delay spread values for the directional case, i.e., for the TX/RX beam combination that provides that highest receive power. 
In case of LOS $86\%$ of the links have RMS-DS within \SI{5}{ns} to \SI{10}{ns}, see Figure \ref{Fig:rms_LOS}, and the median is \SI{8.7}{ns}. For NLOS, the directional RMS-DS vary from \SI{8}{ns} to \SI{70}{ns} as seen in Figure \ref{Fig:rms_NLOS}. We thus see that the ratio of omni-directional to directional delay spread is on the order of 3, a result that is comparable to the results in \cite{Rappaport_et_al_2015_TCom} for urban environments. 

%
 
 \begin{table}[tbp]\centering \caption{Parameters of the RMS-DS} \label{tab:rms}
   \begin{tabular}{|l|c|c|c|c|}\hline
       & Median(ns) & $\mu$ & $\sigma$  & P-value  \\ \hline
     28th St - omni & 25.63 & -7.58 & 0.263 & 0.889\\
     28th St - directional & 8.19 & -8.10 & 0.101 & 0.254\ \\
     NLoS - omni & 67.18 & -7.20 & 0.156 & 0.664\\
     NLoS - directional & 28.71 & -7.55 & 0.271 & 0.991\\  \hline
   \end{tabular}
 \end{table}

\subsection{Extracted Multi-paths}
\begin{figure}[tbp]\centering
  \includegraphics[width=0.96\linewidth, viewport=38 180 570 600, clip=true]{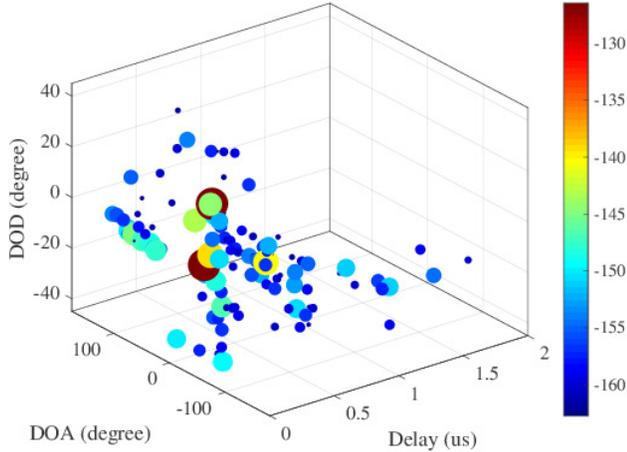}\caption{Extracted multi-path components}\label{Fig:mpc}
\end{figure}

By performing 3-dimensional peak detection in the $P(\theta_{TX},\theta_{RX},\tau)$ we extract the multi-path components (MPC) with the information of; direction of departure (DOD), direction of arrival (DOA), delay and path gain. To avoid the ghost paths due to sidelobes of the beams, for every delay bin, we filter out any MPCs with 10 dB or less path gain compared to the highest power MPC in the same delay bin. The extracted MPCs for the RX\_14 are shown in Figure \ref{Fig:mpc}.

\section{Conclusion} \label{sec:conc}

In this paper we presented results from a channel sounding campaign in a residential suburban environment at 28 GHz. The novel design of the channel sounder allowed phase-coherent measurements of all TX and RX angles. For path-loss, we provided parameters for both ABG and CI models. Although the environment is not urban, we saw that the mean RMS-DS results are inline with the Urban micro-cellular model provided in \cite{3GPP_5G_2016}. We showed that the channel sounder used in this campaign is capable of angular investigations for both TX and RX. In the future, we will provide statistics for angular spreads, perform more measurement campaigns to investigate the outdoor-to-indoor penetration loss and foliage effects.

\section*{Acknowledgement}
Part of this work was supported by grants from the National Science Foundation. The authors would like to thank Sundar Aditya, Vinod Kristem, He Zeng for efforts in helping the measurement campaign and Dimitris Psychoudakis, Thomas Henige, Robert Monroe for their contribution in the development of the channel sounder.

\bibliographystyle{IEEEtran}
\bibliography{mmwave,60ghz_models}

\end{document}